# ALTERNATIVE APPROACH TO 3D DISPLAYING

Oleg G. Semyonov

*State University of New York at Stony Brook, 214 Old Chemistry Building,
Stony Brook 11794, USA*

A method for displaying volumetric images, which exploits our binocular vision and does not require eyewear, is discussed. The display can be rendered as a matrix of pivoting micromirrors irradiated by a light beam; each micromirror focuses its pixel beams to the same point of displayed volumetric image. 3D perception of image can be achieved by scanning the point of beams intersection over a virtual surface of displayed image in space.

OSIS Codes: 120.2040; 110.6880

## 1. Introduction

The dream about a device generating optical volumetric replicas of real objects, just as we see them by eyes, persisted over centuries and recently found its second wind after invention of video imagers and image processors. Image processing has already become a huge scientific area where thousand researchers are actively working. The engineering efforts are mostly concentrated on the 'old good' two-dimensional imaging however a growing team of engineers in three-dimensional (3D) vision area is also working actively. The current "demand for 3D stems more from whim then necessity, which makes it difficult to gauge exactly what kind of image is sought" [1] to be preferable for a potential customer, however the likewise whim seemed to drive, in its time, the engineers toward the prominent progress in automobile industry and, recently, toward the boom in the cellular phones with cameras. A variety of 3D stereo-displays and quasi-volumetric imagers has been proposed and some have already appeared on the market. Nonetheless, the common vision is a display round with people, each seeing what they would see if they were looking at a solid object instead of an image.

Virtually all the tricks for achieving 3D-perception of images, from paintings to modern stereo movies and displays, are based on optical illusion. 2D-images on a flat surface are either distorted in accordance with the rules of perspective to produce the perception of depth, or consist of two images separately for right and left eyes to create a stereo effect. While visual art such as photography and cinematography uses perspective [2] together with light and shade play, interposition, aerial perspective (blurring of distant objects as if they were in haze), textual gradient, and motion parallax, [3] leaving the rest to our imagination in perceiving the depth, stereoscopy mimics, in addition to all the tricks above, our binocular vision to actually see the depth by forcing our brain to arrange the objects of a flat image as if they were hovering in 3D space. However, despite the prominent eye-catching effect of stereoscopy and the persistent efforts to promote the stereoscopic systems, they have not found a proper recognition in mass production so far. There are two reasons for this slow advance: firstly, the stereoscopic systems are costly in comparison with the conventional 2D systems and secondly, they need a kind of eyewear to watch stereoscopic images: on a color-multimplexed (anaglyph) display that creates images of different colors for left and right eyes, on a polarization-multiplexed display producing images of mutually perpendicular polarization, or on a time-multiplexed display where the left and right views are sequentially shown. The recently appeared autostereoscopic displays [1,4] require no eyewear; the views are spatially multiplexed on a pixel-addressable screen and then separated by an optical layer which sends the left view and the right view at different angles to be perceived by left and right eyes separately. All stereoscopic methods are suffering from two drawbacks: firstly, the angle of viewing is limited (as well as the distance to the screen in the case of autostereoscopy) and secondly, at any angle of viewing a viewer sees the same image, i.e. the image does not rotate with the change of angle of view because the viewer sees actually through the viewpoints of the cameras.

One of the first truly volumetric imagers was developed by NASA. [5] The image capabilities of the system stem from the technique that uses a laser-light projection apparatus and a rotating "vortex" screen to create a "viewing box." The system displays nearly 34 million "voxels" (i.e., x, y, and z points corresponding to the pixels of a 2-D image) of data, enabling a user to view a 3-D shape from multiple perspectives without the need for special viewing aids or goggles. The information about another system generating the real spatial image with 360º view angle was published in 2002 [6] and 2004. [7] The system is based on a rotating screen, where a sequence of 5940 2-D images from a XGA-resolution light modulator is projected. This combination of hardware and software generates 3D volumetric images inside a vacuumed transparent spherical dome, which can be seen from a full 360 degrees without goggles. Nevertheless, this masterpiece of optomechanics is far from mass production and out of reach of ordinary consumers.

Our dream is about something like communication devices shown in sci-fi movies, where a 3D image is hovering in air without screens or scattering media being observed from any direction without goggles or other eyewear. A possible approach to displaying the volumetric images which can be seen as hovering in space without reflecting screens or light scattering fog and which can be observed without eyewear is outlined below.

## 2. Perception of image

Our binocular vision is based on the separate perception of light beams emitted by an object at different angles. We see an illuminated object when it scatters light emitted from natural or artificial sources and every point on the object's surface radiates the scattered light more or less isotropically. An eye catches a fraction of scattered light from a particular point on the surface within a solid angle of vision-by-eye $\Omega = A/D^2$, where A is the area of pupil and D is the distance to the object (Fig. 1). The object is seen slightly different by left and right eyes, because our eyes see the object from different angles: the closer is the object, the lager is the stereo-angle $\alpha = 2\arctg(d/2D)$, where d is the distance between the eyes, and the larger is disparity between the images on the retinas of left and right eyes.

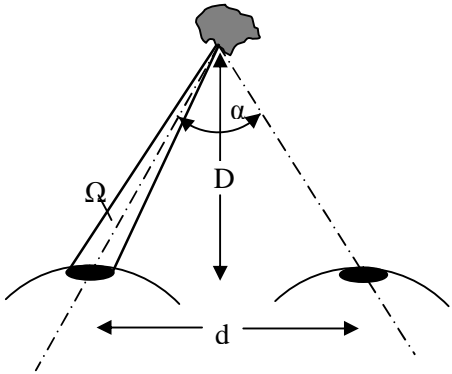

Fig.1. Binocular vision (see details in the text) of 3D-objects: $\Omega$ is the visibility-by-eye cone of a point on the surface of objects, D is the distance between observer and object, d is distance between two observer's pupils, and $\alpha$ is the stereo-angle.

To obtain a volumetric replica of a real object, which can be perceived as a real object, an optical system and/or a multiplexing display is supposed to create a 3D-image in space, which imitates the real object in the sense that every point of image radiates a divergent quasi-isotropic fan of rays having intensity that mimics the intensity distribution of scattered light on the surface of a real object. In principle, such the volumetric image can be produced by a properly designed optical system. For example, an optical toy "Mirage" (Edmund Scientific) consisting of two spherical mirrors creates the object's 'apparition' in air above the toy, which looks indistinguishable from the real object placed near the surface of the bottom mirror. As seen from Fig. 2, the system of mirrors transforms each luminous point of the object's surface to the conjugative point on the virtual 'surface' of the volumetric image located above the upper mirror; the image is rotated by 180º with respect to the real object but conserves its 3D perception when viewed by eyes. A viewer sees the volumetric image because his binocular vision works the same way as it would be if a real object were placed instead of image. One can walk around and watch the image from any side. The 'mirage' effect is achieved because: a) every virtual point of the image is created by the bottom mirror as a converging-in-space fan of rays originated from different points on the mirror surface and, therefore, every point of this image emits outside a divergent fan of rays the same way as if a real object was placed instead of the image and b) left and right eyes of a viewer capture the rays emitted by a particular point of image from two different solid angles of visibility-by-eye $\Omega$, which are actually the continuations of rays reflected from different corresponding areas S on the surface of the bottom mirror located on the lines of sight (shown by the dashed lines in Fig. 2): $S = r^2 \Omega = A r^2/D^2$, where r is the distance between the viewed point of the image and the surface of the mirror along the line of sight and D is the distance between the eye and the observed point of image. The important feature is that one can see image only when the light-reflecting portion of surface of the bottom mirror and the corresponding point of image, where the rays are focused to, lie on the line of sight; it is impossible to see image when the line of sight is inclined above the edge of bottom mirror and therefore it is impossible to see other objects of real world through the mirage image as it is often shown in sci-fi movies. Nevertheless, the idea looks attractive for application in optoelectronic displaying.

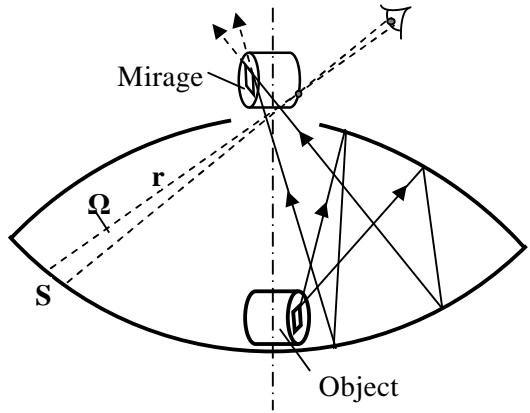

Fig.2. Optical diagram of Mirage toy with the ray paths traced for a couple of rays emitted from a point on the surface of a real object located on the bottom mirror. Here r is the distance between a point on the surface of bottom mirror, where a ray is reflected, and a corresponding point on the image surface, S is the area of bottom mirror participating in perception of a point of image by an eye, and $\Omega$ is the visibility-by-eye angle (dashed lines) of this point.

Imagine a matrix of light-emitting pixels on a spherical surface installed instead of the bottom mirror in Fig. 2 (it can be a matrix of laser diodes or a matrix of micro-mirrors irradiated by a light source), which can be controllably tilted to direct every pixel microbeam in any direction within the desired angle of visibility determined by the display's size, the distance between viewer and display, and the size of image we intend to create. Suppose all pixel beams can be directed to one and the same point in space and that every pixel beam can be focused, if necessary, on this point to produce a point of a spatial image (Fig. 3a). There is no need for mirrors or other additional optics and there is no need for a real object; nonetheless, such the point will be perceived by a viewer as a luminous point located in space being visible when the viewer is in the angle of visibility $\Omega_o$ because the point radiates a divergent fan of rays as if it belong to a real object.

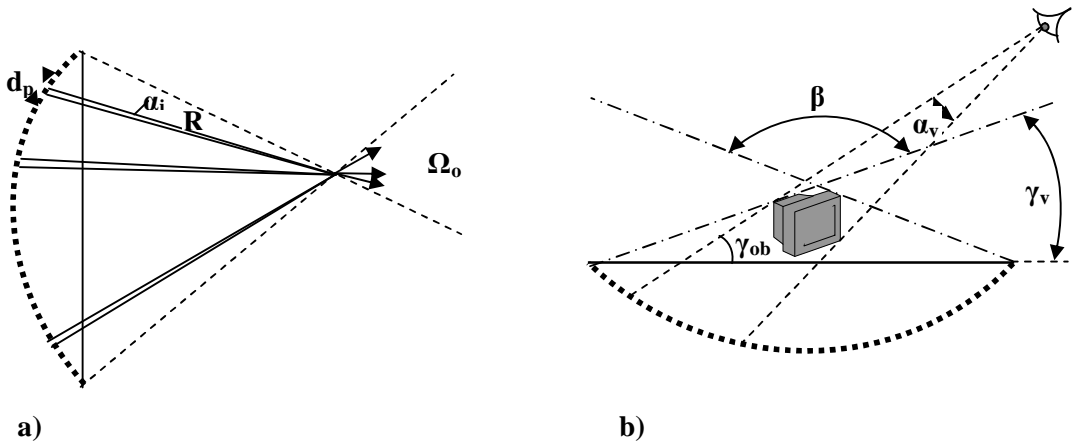

a)             b)

Fig.3. a) Displaying a virtual image of a point in space by a spherical display made of pivoting light-emitting pixels focused on this point; b) displaying a virtual volumetric image of an object: $\Omega_0$ is the solid angle of visibility where a given point can be observed, R is the distance between a current point of beams crossing and an emitting pixel having size $d_p$, $\alpha_i \sim d_p/R$ is the focusing angle of a particular pixel beam, $\alpha_v$ is the total angle of image observed by an observer, and $\beta$ is the total elevation visibility angle of multiplexed volumetric image observed from any direction around a horizontal display.

The rest looks obvious. What is needed is to scan the point of crossing of all pixel beams over the virtual volumetric surface of image, we intend to display, or, more exactly, to scan it over a portion of virtual surface of image on its far side from the origin of a given pixel beam on the display surface (i.e., over a corresponding nearest side of this virtual image with respect to a viewer) within a desired solid angle of visibility. The scans can be rendered analogous to TV-scanning, where the frames are made of horizontal scan lines. To avoid flickering, this scanning must be sufficiently fast, say, 30 to 60 frames per second. The 'apparition' will be perceived volumetrically such as the image of the Mirage toy because, like in the case of a real object, every point of this virtual image emits the divergent quasi-isotropic fan of rays into the visibility angle and our binocular vision works as usual. No image is generated on the display itself; information about the current positions of virtual points of image (about every point of crossing of beams in space) together with the intensities of particular beams at a particular moment and the perceived color can be multiplexed as a time sequenced stream of data addressed to every beam-generating pixel. Ideally, every pixel has to be addressed individually to direct the intensity- multiplexed pixel beams to a current scan point of volumetric image. The image can be viewed from different sides and a 360 degrees view can be achieved if the display is positioned horizontally (Fig. 3b). In the last case, the image is visible when the viewing angle $\alpha_v$ is less then the total angle of visibility $\beta$ and the elevation angle of observation $\gamma_{ob}$ exceeds the elevation angle of visibility $\gamma_v$.

In a real system, the distortions of pixel beams would blur the image. In particular, the image quality can be influenced by inaccuracy of pixel beams alignment, optical diffraction, beam divergence, and optical aberrations.

*Diffraction.* Focusing of pixel beams is preferable when the angle of diffraction of a pixel beam $\alpha_{diff} = 1.2\lambda/d_p$ is less then the focusing angle $\alpha_i = d_p/R$, i.e. if $R < d_p^2/1.2\lambda$, where $R \approx F$ is the distance from a particular portion of display surface to a current point of image (approximately equal to the focal distance F if focusing is produced by each properly formed reflection pixel) provided other factors, such as optical aberrations, are negligible. The halfwidth of image's spot, which is the point of crossing of pixel beams on the condition of ideal alignment, can be estimated as $d_s \approx 0.6\ R\lambda/d_p$ (for example, $d_s \approx 0.1$ mm, if R = 10 cm and the optical aperture of each pixel is equal to 250 μm). The graphs $d_s(R)$ (resolution due to diffraction versus distance from the display surface) are shown in Fig. 4 for various optical sizes of pixels $d_p$. Focusing of pixel beams becomes useless when $R > d_p^2/1.2\lambda$.

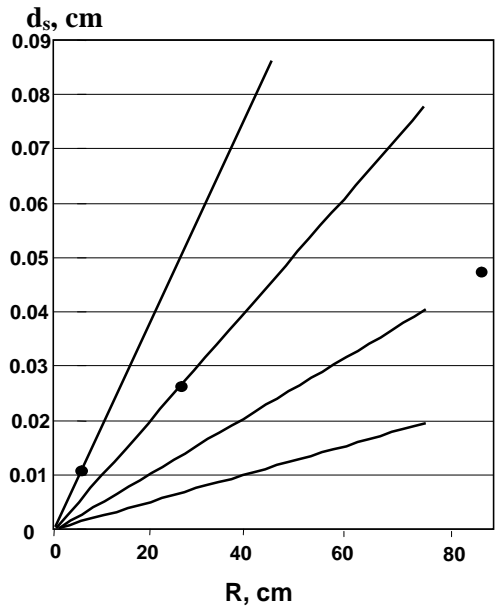

Fig.4. Diffraction spot $d_s$ ($\lambda$ = 500 nm) versus distance R from the display surface for the pixel sizes: $d_p$ = 100 μm (upper line), 250 μm, 500 μm, and 1 mm (bottom line). Solid circles on the lines indicate the distance $R_{eq}$ at which the diffraction spot diameter is equal to pixel size. Additional focusing of pixel beams is advantageous for better resolution if $R < R_{eq}$ and useless otherwise.

*Divergence.* In addition to diffraction, the pixel beams will be more or less divergent by the nature of the real sources of electromagnetic waves (e.g., laser beams consisting of multiple angular modes or light beams formed by optical systems when emitters of finite size are used), so the wave-fronts of pixel beams will never be absolutely flat or precisely spherical. If no focusing is implemented, the cross-sections $d_b$ of unfocused beams at a distance R will be the functions of their divergence half-angle $\gamma$: $d_b = d_p + 2\gamma R$. Since the divergence of multimode commercial lasers $\gamma$ is ~ $10^{-3}$ radians, the images must be generated at the distance $R \leq 0.3 d_p/2\gamma$, if the tolerable increase of cross-section of unfocused beams is taken $0.3 d_p$. In the case of focused pixel beams, the caustic size in the focal plane $d_c = 2\gamma F$, where $F \approx R$ is the focal distance, and the focal distance $F$ and the distance $R$ to the image are limited by the condition $R \approx F \leq d_{c0}/2\gamma$ to obtain the desired resolution $d_{c0}$ (caustic size), for example $R \leq 30$ cm for $2\gamma = 10^{-3}$ when $d_{c0} = 300$ μm.

*Optical aberrations.* When the pixel beams are focused by a sort of optical system, optical aberrations can also worsen the resolution. Approximation of paraxial optics is valid for the small focusing angles $\alpha_i = d_p/F$, when $d_p \ll F$, so only spherical aberrations and chromatic aberrations determine the size of caustic near the focal points of pixel beams and the depth resolution of color images. Chromatic aberration can be zeroed, if the pixels are made as reflective spherical micro-mirrors. Spherical aberration $\delta_{sph} \approx 0.07\, d_p^3/F^2$ is quite small when $d_p < F$ and virtually plays no role in image blurring in comparison with the blur caused by the divergent beams and inaccurate alignment.

*Blur due to uncoordinated pixel beams.* Apparently, the blur due to bad alignment of pixel beams on the scanned spots of virtual image will be determinant in perceiving the image sharpness. It strongly depends on angular accuracy of pixel pivoting mechanisms. It should be noticed, that keeping all pixel beams crossed at a current scan point at every particular moment, as it was outlined above, is not a necessity. The scans of separate pixel beams could be time-uncorrelated to an extent; it would be sufficient to provide spatial accuracy of their scanning along the same scan lines on the surface of virtual image with proper multiplexing of brightness, color, and focal distance (if necessary), when a particular beam cross-sects a particular point on the virtual surface of displayed image.

### 3. Technical problems and perspectives

The "Mirage"-like method of volumetric imaging requires a display consisting of multiple pivoting micro-sources of light beams with controlled tilt to synchronously direct each pixel beam to every point of displayed image without deteriorating the desired resolution. In particular, it can be based on the MOEMS technology developed for the reflecting microdisplays equipped with controllably tilting micro-mirrors. Texas Instruments, for example, manufactures the microdisplays consisting of arrays of torsion micro-mirrors. The limiting factor in application of this technology for the proposed system is that the tilt of mirrors in one direction is actually uncorrelated oscillations. The technique could be implemented, if the angular positions of the oscillating mirror were known precisely at every moment to properly multiplex the brightness and color of each pixel beam as functions of tilt because, as it has been mentioned above, the tilts of microbeams along the scan lines could be time uncorrelated while their raster in the perpendicular direction must be coordinated and accurate. Individual addressing of every pixel is needed in this case. Another way is to make a display consisting of pixel mirrors synchronously tilting by sufficiently large angles (tens degrees) both in azimuthal and polar directions with accuracy not worse then $\delta\varphi = d_b/R$, where $d_b$ is the cross-section of pixel beam (or the diameter of caustic) at the distance R, to provide a total blur of multiplexed points of image not exceeding significantly the blur from a single pixel beam. The existing MEMS devices developed by MEMs Optical and Applied MEMs companies with two-axis tilt execute the maximum angles of tilt of 500-μm mirrors from to 2 to 6 degrees, which is too small for the considered application. Quite promising could be the scanning micro-mirror MEMS technology developed by Microvision. [9, 10] Their 1-mm$^2$ mirrors can be controllably tilted in both directions by the angles up to 30º. The disadvantage is big size of a die determined by a bulk magnetic system used for tilt control. Yet another approach to control the tilt is to implement thermomechanical actuators in combination with electrostatic actuators to control the dual-servo mirrors. [11 - 15] As for intensity multiplexing, the use of Grating Light Valve (GLV) [16] technology with the beam intensity controlled by a grid of ribbons of controllable displacement could be also thinkable, when used in combination with the tilting mechanisms to allow both scanning and intensity multiplexing of each pixel. The drawback of GLV technology is that each pixel actually produces several beams (at least, one zero order and two first order beams) and parasitic images can be perceived at different angles of observation.

The high fill-factor arrays of micromirrors with electrostatically controlled and individually addressed 2D-tilt have been reported. [17] Providing the problem of synchronous tilt of micro-mirrors with alignment of all pixel beams to one scanned spot of image is solved, the display can be illuminated by a large-aperture light beam with flat or converging wavefront (even a fanned beam can be used, if the spherical micro-mirrors have the compensating focal distances). The advantage is that it is not necessary to address a particular pixel for brightness and color multiplexing; instead, the total illuminating beam can be color-multiplexed as a time sequence of intensities of composite colors addressed to a current point of image, if the beam consists of, for example, three basic color beams.

Adjustable and addressed focusing of each pixel beam for every transient point of image would be a serious problem. In the case of individually addressed pixel mirrors, every mirror should have adjustable focal distance using the adaptive optics technology. In practice, it seems to be sufficient to control the focal distances of pixel beams originated from relatively small areas of display by locally changing the form of wavefront of illuminating beam with a kind of adaptive optics. Another possibility is to make the display itself consisting of micromirrors with the piston-like movement in addition to tilt using the adaptive optics approach. [18] In the case of relatively small depth of images, all the pixels could be made of the same form with the same focal distance equal to the average distance of virtual image from the display surface, and the wavefront distortion of illuminating light beam by the adaptive optics would be sufficient to achieve the necessary control of focal distance and the uniform sharpness over the total image.

Small displays like the screens of cellular phones can be made flat without significant deterioration of resolution of images created within several centimeters from the screen. A spherical form is preferable for larger displays, when the images are maid by the focused pixel beams; firstly, the difference in R from the pixels at opposing edges of a display to the portions of images that are out of the axis of symmetry can be minimized and secondly, the maximum angles of tilt $\theta \approx \mathrm{arctg}(d_i/R)$ of all pixel mirrors can be kept sufficiently small, where $d_i$ is the transverse size of displayed images, while providing a broad angle of total visibility β. For displaying the extended images, say, a chess-board or a virtual battlefield, it would be more expedient to divide a flat display into several areas by means of corresponding software, each displaying a portion of the total image in a given direction determined by the maximum angle of tilt θ and the total elevation angle of visibility β. In the last case the display can be made of any customized form.

Displaying the fictional volumetric objects, such as cartoon characters, molecular structures, object of battlefield, etc., is mainly the task of corresponding software. Displaying the real objects would require shooting of a scene by a multi-camera system, for example, by the system described in [5] with the followed reconstruction of 3D images using the proper software to transform the multi-camera flat images to the stream of data analogous to the system described by Park and Inoue. [8]

One of anticipated software problems will be, in particular, the "inverse occlusion", when a portion of volumetric surface visible from one position of a viewer must be made invisible from another direction of viewing (Fig. 5). A possible solution is to scan the corresponding pixel beams only over the farthest portions of surface from the given pixel in the case when two or more image surfaces intersect the line of sight between an eye and a given pixel, i.e.

to extinguish the beams from the pixels that 'see' more then one surface when the scanned point of the beams crossing is on a surface, which is supposed to be invisible for a given direction of displaying. On the other hand, for many applications all the surfaces could be made visible through each other including cavities and inclusions which could also be contrasted by artificial colors.

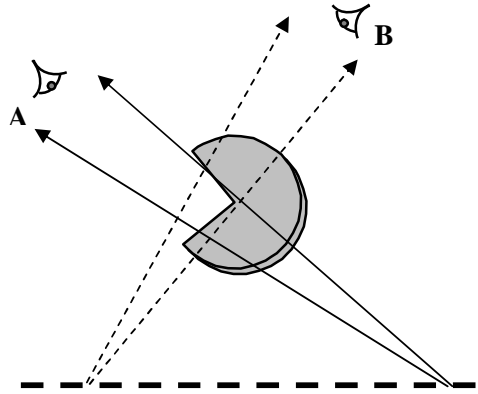

Fig.5. Phenomenon of "inversed occlusion." The "cut-out" portion must be made visible for a viewer A, but invisible for a viewer B.

Obviously, the beam-emitting pixels cannot cover the display surface entirely because the dies are normally larger then their optical apertures. The light-emitting surface of displays will look rather as a polka-dot texture with the light-emitting spots on dark background. To see the image without the blind spots, density of light-emitting dots must be sufficient to ensure that at least one emitting pixel falls into the cone of visibility-by-eye $\Omega$ (Fig. 2) for any point of virtual image and for any angular position of a viewer within the total visibility angle. It means that the distance between the neighboring pixels $d_d$ should not exceed $Rd_{pup}/D \approx 0.4\ R/D$ cm, where $d_{pup} \approx 4$ mm is the average pupil diameter. For example, $d_d$ must be less then 400 μm for an image located at a distance of 3 cm from the display surface, being observed from the distance $D = 30$ cm. Apparently, this condition is not absolute: we are used to slightly patchy and/or blurry images of conventional displays, so some spottiness can be tolerable.

## 4. Conclusion

It seems viable to create true 3D-volumetric images with a display made of tilting micro-mechanical pixels that emit quasi-parallel or focused light beams of multiplexed intensity and color, so all the beams are crossed at a transient point that can be controllably scanned over the surface of virtual image. The optical blur due to diffraction, micro-beam divergence, and optical aberrations can be kept sufficiently small to obtain satisfactory resolution at reasonable distances from the display and for relatively wide range of viewing distances. The main hardware problem is to manufacture the displays of sufficiently dense texture of micro-mechanical pixels able to execute the controllable two-axis tilt of pixel beams within the sufficiently large azimuth and polar angles and to precisely scan the point of crossing of micro-beams over a virtual volumetric surface of image with a flicker frequency exceeding 30 frames per second. The pixel-addressed focusing of each beam on the surface of virtual image can be implemented, if necessary, using the adaptive optics technique in combination with intensity and color multiplexing. Provided the problems of micro-mirrors synchronous tilting and micro-beams alignment to a transient scanned point are solved, the illuminating light beam as a whole can be intensity, color, and focal distance multiplexed, being addressed to a particular spot of displayed image. The anticipated software problem is to provide the color and brightness multiplexing coordinated with the tilt control of micro-mirrors and the wavefront control by the adaptive optics. Currently, there are no displays consisting of arrays of two-axis reflecting pixels with controllable tilt, however the technology of dual-servo micromirrors is developing rapidly and the manufacturability of such displays seems to be technically achievable.

**References**

1. A. Travis, "Autostereoscopic 3-D display," Applied Optics, **29**, 4341 (1990).

2. W. F. Powell, *Perspective (Artist's Library series #13)*, (Walter Foster Publishing, 1990)

3. L. Lipton, *StereoGraphics Developer's Handbook*, (StereoGraphics Corporation, 1991)

4. S. Pastoor and M. Wöpking, "3-D displays: A review of current technologies," Displays, **17**, 100-110 (1997).

5. "Speaking Volumes about 3-D," http://www.sti.nasa.gov/tto/spinoff2002/ps_6.html

6. P. Hill, "SID News," Laser, optics and photonics researches and news, http://optics.org/articles/news/8/5/26/1#3ddisplays

7. "Perspectra Spatial 3-D System," Photonics Spectra, Photonics Awards, 2004

8. J. Park and S. Inoue, "Arbitrary view generation using multiple cameras, " in *Proceedings of International Conference on Image Process*, v. 1, 149-153 (1997).

9. H. Urey, "Retinal Scanning Displays," in *Encyclopedia of Optical Engineering*, (Marcel-Dekker, 2003), 2445-2457.

10. J. Yan, S. Luanava, V. Casasanta, "Magnetic Actuation for MEMS Scanners for Retinal Scanning Displays," in *MOEMS Display and Imaging Systems*, Proc. SPIE **4985**, ISBN 0-81940-4785-4, 115-120 (2003).

11. K. Wang, K. F. Böhringer, M. Sinclair, "Low Voltage and Pull-in state Adjustable Dual Servo Scanning Mirror," in 5$^{th}$ Pacific Rim Conference on Lasers and Electro-Optics (CLEO/Pacific Rim 2003)

12. O. Tsiboi, Y. Mizuno, N. Kona, H. Soneda, H. Okuda, S. Ueda, I, Sawaki, and F. Yamagishi, "A rotational comb-driven micromirror with a large deflection angle and low drive voltage," in *Proceedings of the IEEE on MEMS*, (Institute of Electrical and Electronics Engineers, New York, 2002), 532-535.

13. A. Jain, A. Kopa, Y. Pan, G. K. Fedder, "A Two-Axis Electrothermal Micromirror for Endoscopic Optical Coherence Tomography," IEEE Journal of Selected Topics in Quantum Electronics, **10**(3), 636 (2004).

14. A. Jain, T. Xie, Y. Pan, G. K. Fedder, and H. Xie, "Two-axis Electrothermal SCS Micromirror for Biomedical Imaging," in Optical MEMS, IEEE/LEOS International Conference on Optical MEMS, p. 14-15 (2003).

15. U. Krishnamoorthy, K. Li, K. Yu, D. Lee, J. P. Heritege, and O. Solgaart, "Dual-mode micromirrors for optical phased array applications," Sensors and Actuators, **97-98**, 21-26 (2001)

16. R. W. Corrigan, D. T. Amm, and C. S. Gudeman, "Grating Light Valve Technology for Projection Displays," in Proceedings of the International Display Workshop, #LAD5-1 (1998)

17. V. Milanović, G. Matus, and D. T. McCormick, "Tip-tilt-Piston Actuators for High Fill-Factor Micromirror Arrays" in the Proceedings of Hilton Head 2004 Solid State Sensor, Actuator, and Microsystems Workshop, Hilton Head (2004).

18. T. G. Biffano, J. Perreault, R.K. Mali, M. N. Horenstein, "Micromechanical Deformable Mirrors", Journal of Selected Topics in Quantum Electronics, **5**, 83-90 (1999)